%% file: primo.tex
\documentclass[preprint2]{emulateapj}
\usepackage{natbib,amsmath,verbatim}
\received{2011 September 26}
\accepted{2011 December 23}
\newif\ifbwfig
\bwfigtrue

\def\eg{{e.g.}}
\def\ie{{i.e.}}
\def\etc{{etc.}}

\def\wo{\ensuremath{w_0}}
\def\wa{\ensuremath{w_{\rm a}}}

\def\LCDM{$\Lambda$CDM}

\def\Av{\mbox{$A_V$}}

\newcommand{\primo}{SN~Primo}
\newcommand{\CCSN}{CC\,SN}
\newcommand{\CCSNe}{CC\,SNe}
\newcommand{\SNIa}{SN\,Ia}
\newcommand{\SNeIa}{SNe\,Ia}

\newcommand{\dmfifteen}{\ensuremath{\Delta\mbox{m}_{15}}}

\def\Hubble{{\it Hubble}}

\newcommand{\JHU}{Department of Physics and Astronomy, The Johns Hopkins University, Baltimore, MD 21218, USA}
\newcommand{\STScI}{Space Telescope Science Institute, Baltimore, MD 21218.}
\newcommand{\Berkeley}{Department of Astronomy, University of California, Berkeley, CA 94720-3411, USA}
\newcommand{\Riverside}{Department of Physics and Astronomy, University of California, Riverside, CA 92521, USA}
\newcommand{\WKU}{Department of Physics, Western Kentucky University, Bowling Green, KY 42101, USA}
\newcommand{\Copenhagen}{Dark Cosmology Centre, Niels Bohr Institute, University of Copenhagen, Juliane Maries Vej 30, DK-2100 Copenhagen, Denmark}
\newcommand{\Arizona}{Department of Astronomy, University of Arizona, Tucson, AZ 85721, USA}
\newcommand{\SantaCruz}{Department of Astronomy and Astrophysics, University of California, Santa Cruz, CA 92064, USA}
\newcommand{\NotreDame}{Department of Physics, University of Notre Dame, Notre Dame, IN 46556, USA}
\newcommand{\TelAviv}{Department of Astrophysics, Tel Aviv University, 69978 Tel Aviv, Israel}
\newcommand{\Rutgers}{Department of Physics and Astronomy, Rutgers, The State University of New Jersey, Piscataway, NJ 08854, USA}
\newcommand{\CfA}{Harvard/Smithsonian Center for Astrophysics, Cambridge, MA 02138.}
\newcommand{\Minnesota}{Department of Astronomy, University of Minnesota, 116 Church Street SE, Minneapolis, MN 55455, USA}

\ifx\pdfoutput\undefined
  \pdffalse
  \DeclareGraphicsExtensions{.eps,.ps}
\else
  \ifnum\pdfoutput=1
    \DeclareGraphicsExtensions{.pdf,.png,.jpg}
  \else
    \DeclareGraphicsExtensions{.eps,.ps}
  \fi
\fi

\newenvironment{inlinefigure}{
\def\@captype{figure}
\noindent\begin{minipage}{0.999\linewidth}\begin{center}}
{\end{center}\end{minipage}\smallskip}

\newcommand{\insertfigwide}[2] {
\begin{figure*}
\begin{center}
\resizebox{\textwidth}{!}{\includegraphics{{#1}}}
\caption{{#2}}
\end{center}
\end{figure*}
}

\newcommand{\insertfigfloat}[2] {
\begin{figure}
\begin{center}
\resizebox{\columnwidth}{!}{\includegraphics{{#1}}}
\figcaption{{#2}}
\end{center}
\end{figure}
}

\newcommand{\figdiscovery}{
(a) Composite image of the \primo\ host galaxy and surroundings in the UDF, from pre-explosion imaging.
(b) CANDELS search-epoch image in F160W (H band), from 2010 October 10.
(c) F160W difference image, showing \primo\ near peak brightness.
\label{fig:discovery}}

\newcommand{\figcolormag}{
Color-magnitude diagram showing the F350LP$-$F160W color diagnostic,
used in the initial identification of \primo\ as a high-$z$ \SNIa\
candidate.  The SNANA light curve simulator \citep{Kessler:2009a} was
used to generate 10,000 SNe at $z = 1.55$, with colors and magnitudes
measured at times between 5 days before and 10 days after peak
brightness, in the observer frame.  Relative frequencies for each
subclass were drawn from \citet{Li:2011} (24\% Ia, 19\% Ibc, 57\% II)
and luminosity functions follow
\citet{Richardson:2002}, \citet{Kiewe:2012}, and \citet{Drout:2011}.
The shaded contours and corresponding histograms show the full extent
of each SN subclass (\ie, they contain 100\% of the simulated objects
for each class).  Type II SNe are shown in blue, Type Ibc in green,
and Type Ia in red.  The solid red line in the central figure
demarcates the region containing $68\%$ of the simulated \SNeIa.
The \SNIa\ simulation uses a parameterized model based on the SALT2
light curve fitter \citep{Guy:2007}.  \CCSNe\ are simulated using 44
spectrophotometric templates that are based on well-observed
low-$z$ \CCSNe\ (hence the appearance of stripes and gaps in
the \CCSN\ contours).
\label{fig:colormag} }

\newcommand{\figdeltadelta}{
The light curve width parameter \dmfifteen\ for the same simulated
SNe shown in Fig.~\ref{fig:colormag}.  The \dmfifteen\ parameter
measures the increase in observed magnitude from the peak to 15
rest-frame days after peak, and is tightly correlated with absolute
magnitude for \SNeIa\ \citep{Phillips:1993}. For each simulated SN we
measure \dmfifteen\ in both F125W ($J$) and F160W ($H$), and plot
contours containing the entire sample in the main panel (blue for Type
II, green for Type Ibc, and red for Type Ia).  Histograms of the 
one-dimensional projections are shown in the top and side panels.  As in
Fig.~\ref{fig:colormag}, the \CCSN\ contours show spotty coverage of
the parameter space, because they are based on 44 discrete templates,
rather than a single parameterized light curve model as used for
the \SNeIa.
\label{fig:deltadelta} }

\newcommand{\figlightcurve}{
  The \primo\ light curve for F125W ($J$) as blue circles and F160W
  ($H$) as red squares. At a redshift of $z=1.55$ these correspond to
  rest-frame V and R bands, respectively.  Solid lines show the
  best-fitting template, as determined by the SOFT light curve
  classification program, and dashed lines indicate the model
  uncertainty \citep[see][]{Rodney:2009}.  The best-fit template is
  based on the light curve of the normal Type Ia SN
  2005cf \citep{WangX:2009}.
\label{fig:lightcurve} }

\newcommand{\figspectrum}{
  The HST G141 grism spectrum of \primo.  The left side depicts the
  spectral data reduction process: the bottom panel shows the 2D grism
  spectrum, the center panel shows the host-galaxy spectrum, smoothed
  and shifted as described in the text, and the top panel shows the
  host-subtracted SN spectrum.  Grey lines show the unbinned spectrum
  in rest wavelength for the known redshift $z=1.55$. Solid points
  show the mean values in 80 \AA\ bins. On the right side, the same
  binned points are shown in each of the three panels, with three
  template spectra overlaid as solid lines for SNe of Type Ia, Ibc,
  and II.  All templates are depicted for the known age of the SN at
  the time of the grism observation: 6 rest-frame days past peak
  brightness.  The vertical grey bands indicate regions where the SN
  spectrum was contaminated by bright emission lines from the host
  galaxy: H$\beta$ and [O~III] on the blue side, and H$\alpha$ on the
  red side.  \label{fig:spectrum} }

\newcommand{\fighubblepoints}{
SN\,Ia Hubble residuals diagram, plotting distance modulus relative to
the \LCDM\ cosmology versus redshift.  Colored points with error bars
show the compilation of $\sim 500$ \SNeIa\ from \citet{Conley:2011}: the
low-$z$ sample in purple and the mid-$z$ SNe in blue and green, with
large open circles showing the mean values in redshift bins of width
0.2. For the high-$z$ range, the ``Gold'' and ``Silver'' \SNIa\
samples from \citet{Riess:2007} and \citet{Riess:2004a} are shown in
gold and grey points, respectively, with open symbols indicating
objects that lack a spectroscopic classification. Separately
normalized histograms along the lower edge show the distribution of
points for each survey, and the red bar shows the expected reach of
the {\it HST} MCT survey presented in this work.
\label{fig:hubblepoints} }

\usepackage{amssymb}  
\usepackage{url}
\usepackage[usenames]{color}  

\shorttitle{\SNIa\ at $z=1.55$ from CANDELS}
\shortauthors{Rodney et al.}
\slugcomment{ApJ, in press} 

\begin{document}

\title{A TYPE Ia SUPERNOVA AT REDSHIFT 1.55 \\ IN HUBBLE SPACE TELESCOPE INFRARED OBSERVATIONS FROM CANDELS}

    \author{
      Steven~A.~Rodney\altaffilmark{1},
      Adam~G.~Riess\altaffilmark{1,}\altaffilmark{2},
      Tomas~Dahlen\altaffilmark{2},
      Louis-Gregory~Strolger\altaffilmark{3},\\
      Henry C.~Ferguson\altaffilmark{2},
      Jens~Hjorth\altaffilmark{4},
      Teddy~F.~Frederiksen\altaffilmark{4},
      Benjamin~J.~Weiner\altaffilmark{5},
      Bahram~Mobasher\altaffilmark{6},
      Stefano~Casertano\altaffilmark{2},
      David~ O.~Jones\altaffilmark{1},
      Peter Challis\altaffilmark{7},
      S.~M.~Faber\altaffilmark{8},
      Alexei V. Filippenko\altaffilmark{9},
      Peter Garnavich\altaffilmark{10},
      Or Graur\altaffilmark{11},
      Norman A.~Grogin\altaffilmark{2},
      Brian Hayden\altaffilmark{10},
      Saurabh W.~Jha\altaffilmark{12},
      Robert P.~Kirshner\altaffilmark{7},  
      Dale Kocevski\altaffilmark{8},
      Anton Koekemoer\altaffilmark{2},
      Curtis McCully\altaffilmark{12},
      Brandon Patel\altaffilmark{12},
      Abhijith Rajan\altaffilmark{2}, and
      Claudia Scarlata\altaffilmark{13}
    }
    \altaffiltext{1}{\JHU}
    \altaffiltext{2}{\STScI}
    \altaffiltext{3}{\WKU}
    \altaffiltext{4}{\Copenhagen}
    \altaffiltext{5}{\Arizona}
    \altaffiltext{6}{\Riverside}
    \altaffiltext{7}{\CfA}
    \altaffiltext{8}{\SantaCruz}
    \altaffiltext{9}{\Berkeley}
    \altaffiltext{10}{\NotreDame}
    \altaffiltext{11}{\TelAviv}
    \altaffiltext{12}{\Rutgers}
    \altaffiltext{13}{\Minnesota}

\begin{abstract}
We report the discovery of a Type Ia supernova (\SNIa) at redshift
$z = 1.55$ with the infrared detector of the Wide Field Camera 3 (WFC3-IR)
on the {\it Hubble Space Telescope (HST)}.  This object was discovered in
CANDELS imaging data of the Hubble Ultra Deep Field, and followed as
part of the CANDELS+CLASH Supernova project, comprising the SN
search components from those two {\it HST} multi-cycle treasury
programs.  This is the highest redshift \SNIa\ with direct
spectroscopic evidence for classification.  It is also the
first \SNIa\ at $z > 1$ found and followed in the infrared, providing a full
light curve in rest-frame optical bands.  The classification and redshift are
securely defined from a combination of multi-band and multi-epoch
photometry of the SN, ground-based spectroscopy of the host galaxy,
and WFC3-IR grism spectroscopy of both the SN and host.  This object is
the first of a projected sample at $z > 1.5$ that will be discovered by
the CANDELS and CLASH programs.  The full CANDELS+CLASH SN~Ia sample will
enable unique tests for evolutionary effects that could arise due to
differences in \SNIa\ progenitor systems as a function of redshift.
This high-$z$ sample will also allow measurement of the \SNIa\ rate
out to $z \approx 2$, providing a complementary constraint on \SNIa\
progenitor models.

\end{abstract}


\section{Introduction}
\label{sec:introduction}


  \insertfigwide{FIG/fig1}{\fighubblepoints}

In their use as ``standardizable'' candles, Type Ia supernovae (\SNeIa) have
become one of the pillars of modern observational cosmology. \SNeIa\
provided the first direct evidence for an accelerating expansion of
the universe \citep{Riess:1998,Perlmutter:1999}, an effect now
commonly attributed to ``dark energy.''  The present challenge for \SNIa\ 
cosmology is to understand the dark energy equation-of-state parameter, 
$w = P/(\rho c^2)$ \citep{Turner:1997,Caldwell:1998,Garnavich:1998b}.
Recent surveys have targeted \SNeIa\ in one of three broad redshift ($z$)
categories, illustrated in Figure~\ref{fig:hubblepoints} as follows. 

\begin{enumerate}
\item{\em Low-$z$.} \SNeIa\ in the low-redshift regime ($0.02 \lesssim
z \lesssim 0.1$) provide the anchor for the \SNIa\ Hubble diagram
\citep{Jha:2006,Hicken:2009,Contreras:2010}.  
These ``local'' SNe have been used to develop empirical tools for
using light curve shapes to classify high-$z$ SNe and measure their
luminosities.

\item{\em Mid-$z$.} At intermediate redshifts ($0.1 \lesssim z \lesssim 1$)
ground-based surveys such as ESSENCE, SDSS, and SNLS\footnote{ESSENCE:
Equation of State: SupErNovae trace Cosmic Expansion; SDSS: Sloan
Digital Sky Survey; SNLS: SuperNova Legacy Survey.}  have built up
samples of hundreds of \SNeIa, testing models that assume a constant
dark energy equation-of-state parameter, \wo\ 
\citep{Wood-Vasey:2007,Kessler:2009,Sullivan:2011}.
Current and future programs such as PTF, Pan-STARRS, DES, and
LSST\footnote{PTF: the Palomar Transient Factory; Pan-STARRS: the
Panoramic Survey Telescope And Rapid Response System; DES: The Dark
Energy Survey; LSST: the Large Synoptic Survey Telescope.} will
provide samples with thousands of \SNeIa\ to $z \approx 0.8$.

\item{\em High-$z$.} The high-redshift range ($0.8 \lesssim z \lesssim
1.5$) has been populated almost exclusively by the {\it Hubble Space
Telescope (HST)} with optical surveys using the Advanced Camera for
Surveys (ACS) \citep{Riess:2004a,Riess:2007,Suzuki:2011}.  These
high-redshift objects reach back to the era of deceleration, enabling
tests of models with a time-varying dark-energy component $w(z)$ and
checks against extreme \SNIa\ systematics.  From the set of
high-$z$ \SNeIa\ with measured light curves, the highest redshift on
record is SN 1997ff at $z \approx 1.7$ \citep{Riess:2001}. This object
was found in a passive host with an old stellar population, strongly
suggesting it is a \SNIa.  However, there is no reliable spectroscopic
measurement of the SN, it has only a sparsely observed light curve,
and the host redshift relies on a photo-$z$ and a questionable
single-line detection. More and better observations are clearly needed
before any inferences about the high-$z$ \SNIa\ population can be
drawn.
\end{enumerate}

Collectively, the \SNIa\ samples out to $z \approx 1.5$ are consistent 
with a description of dark energy as the cosmological constant,
$w(z)=-1$ \citep{Riess:2007,Hicken:2009,Sullivan:2011,Suzuki:2011}.
Figure~\ref{fig:hubblepoints} shows a recent collection of
$\sim 500$ \SNeIa\ from \citet{Conley:2011}, with distances plotted
relative to the best-fit \LCDM\ cosmology. Histograms on the lower
edge show the redshift range of each contributing survey.  With the
addition of the Wide Field Camera 3 infrared detector (WFC3-IR) on
{\it HST}, a new window has been opened, allowing the detection of \SNeIa\ 
at $z>1.5$.  This very high redshift regime provides an excellent
laboratory in which to test for possible evolution of the \SNIa\
population \citep{Riess:2006}.  

The ratio of dark energy to matter density (now $\sim 2.7$) decreases
with redshift as $\rho_\Lambda/\rho_M\propto (1+z)^{3w} \propto
(1+z)^{-3}$ \citep[\eg,][]{Turner:1997}, so the $z>1.5$ universe is
matter dominated.  Parameterizing the dark energy equation of state as
$w=\wo+\wa(1-a)$, current observations find $\wo=-1\pm0.2$ and
$\wa=-1\pm1$ \citep[\eg,][]{Sullivan:2011}.  Changes in \wo\ and \wa\
consistent with these constraints would affect the observed \SNIa\
magnitudes at $z>1.5$ by less than 0.1 mag.  This means that a larger
deviation in the peak magnitudes of high-$z$ SNe would provide
evidence for evolution of the SN~Ia population.

\citet{Riess:2006} considered \SNIa\ progenitor models that predict 
a decrease in the observed \SNIa\ luminosity for objects with a higher
initial progenitor mass, due to changes in the internal C/O ratio at
the time of explosion \citep{Dominguez:2001,Hoeflich:1998}.  If such
an effect exists, then we might expect to see its signature becoming
apparent in the \SNIa\ population at $z>1.5$: the universe is $<$4 Gyr
old at these redshifts, so low-mass stars are still on the main
sequence, and thus the \SNIa\ progenitor stars must necessarily be
more massive.  

  \insertfigwide{FIG/fig2}{\figdiscovery}

The high-$z$ \SNIa\ sample also provides an important constraint on
progenitor models through the measurement of \SNIa\ rates.  Binary
stellar population synthesis combined with models of \SNIa\ explosion
conditions can provide a prediction for the delay-time distribution
(DTD) that should be observed for the \SNIa\ progenitor population at
any redshift. Convolving this predicted DTD with measurements of the
cosmic star formation rate produces a prediction for the \SNIa\ rate
as a function of redshift.  There is general consensus on the
measured \SNIa\ rate out to $z \approx 1$, and these observations can
be well matched by a number of progenitor
models \citep[see, \eg,][hereafter G11, for a recent
compilation]{Graur:2011}. It is at $z>1$ that the \SNIa\ rate starts
to be strongly dependent on the shape of the DTD, and the sparse 
measurements in this regime place only modest constraints
on possible progenitor models
(\citealp{Strolger:2004}; \citealp{Kuznetsova:2008}; \citealp[hereafter
D08]{Dahlen:2008}; G11).  Our  WFC3-IR survey will provide a
significant improvement in the high-$z$ \SNIa\ rate measurements by
substantially increasing the spectroscopically confirmed \SNIa\ sample
at $z>1$.

Metallicity effects should be more pronounced in the $z>1.5$ \SNIa\
sample, and may be apparent in the observed \SNIa\ rates.  Consider
a \SNIa\ progenitor system with a long delay time of $\sim 3$ Gyr. For
this object to be observed at $z \approx 2$, the formation redshift would
have been $z \approx 10$, close to the redshift of
reionization \citep{Komatsu:2011} when the universe was {\em only}
$\sim 300$ Myr old.  This long-delay SN would have been born in the
first generation of stars, in an extremely metal-poor environment.
The progenitor model of \citet{Hachisu:1996} invokes a
wind from the accreting white dwarf, and therefore requires a high
metallicity as a prerequisite for \SNIa\ explosion.  If this
progenitor track accounts for a significant fraction of the \SNIa\
population, then the \citet{Hachisu:1996} model would suggest that the
observed \SNIa\ rate may be much lower at $z>1.5$ than otherwise
expected.

\input{lightcurve}

In this work we present an early result from the CANDELS+CLASH
Supernova project (PI: Riess).  This search-and-follow SN program is a
composite survey combining the SN search components from two MCT
programs, both of which were designed with cadence and filter choices
that enable the detection of high-z SNe using the WFC3-IR detector.
The Cosmic Assembly Near-infrared Deep Extragalactic Legacy Survey
(CANDELS, PI: Faber \& Ferguson) is a wide-field survey targeting five
famous fields (GOODS-S, GOODS-N, COSMOS, EGS, and UDS). Observations
and data processing for CANDELS are described by \citet{Grogin:2011}
and \citet{Koekemoer:2011}, respectively.  The second MCT program is
the Cluster Lensing And Supernova survey with Hubble (CLASH,
PI: Postman), which is targeting 25 low-redshift galaxy
clusters \citep{Postman:2011}. The SNe contributing to the
high-redshift \SNIa\ sample from CLASH will be located behind the
clusters in the primary fields, or in the extended survey area
provided by {\it HST} parallel observations.  Both programs will span three
years, from 2010 to 2013.

Within the first epoch of CANDELS imaging, the SN team discovered a
high-z SN candidate, dubbed \primo.  As described below, this object
was later confirmed as a \SNIa\ at $z=1.55$, making it the highest
redshift \SNIa\ with a spectroscopic confirmation.  The detection and
follow-up observations of \primo\ are detailed
in \S\ref{sec:discovery}-\ref{sec:grism}, and
in \S\ref{sec:discussion} we explore the potential for this survey to
extend the \SNIa\ sample to $z \approx 2$.

  \insertfigfloat{FIG/fig3}{\figcolormag}
  \insertfigfloat{FIG/fig4}{\figdeltadelta}

\section{Discovery and Follow-up}
\label{sec:discovery}

\primo\ was found in CANDELS search images of the GOODS-S field
collected on 2010 October 10 (UT dates are used throughout this
paper). It was detected in both WFC3-IR search
filters (F125W = $J$ and F160W = $H$) as well as the broad ``white light''
filter of WFC3-UVIS (F350LP = $W$); see \citet{Grogin:2011} for a
complete description of the CANDELS observations. The use of F350LP to
discriminate core-collapse SNe (\CCSNe) from \SNeIa\ is presented
in \S\ref{sec:color}.  It was located within the 11.5 arcmin$^2$ Hubble
Ultra Deep Field (HUDF) region, which had been recently observed
using WFC3-IR under a Cycle 17 {\it HST} program (GO-11563,
PI: Illingworth).  \primo\ was found in a search by eye of difference
images that were constructed by subtracting a deep template HUDF image
from the October 10 CANDELS images, as shown in
Figure~\ref{fig:discovery}.

After discovery in the CANDELS imaging data, \primo\ was recovered in
prior F160W observations from {\it HST} program GO-11563 taken in
September, 2010 (see Table~\ref{tab:lightcurve}).  Using the
non-detection on UT Sep 02.9 and the clear detection on Sept 14.9 as a
guide, we made an initial estimate that the explosion occurred between
UT Aug 27 and Sep 14.  This would make \primo\ 26-45 observer-frame
days past explosion at the time of discovery on Oct 10.9. 

The host-galaxy photo-$z$ was estimated at $z_{\rm phot}=1.56$
(\S\ref{sec:host}), meaning that the SN was 10-18 rest-frame days past
explosion. If \primo\ was a normal SN from one of the primary
sub-classes, this indicated that it was close to peak brightness.
Comparing the magnitudes and colors of \primo\ to simulations based on
low-$z$ templates, \primo\ was found to be consistent with a \SNIa\
near peak at $z>1$ (\S\ref{sec:color}).

These observations gave us an early indication that this object was
very likely to be a \SNIa, so we triggered target of opportunity
follow-up observations from the ground with the Very Large Telescope
(VLT) and from space with {\it HST}.  The VLT observations
(Frederiksen et al. 2012, in preparation) revealed a spectroscopic
redshift of the host galaxy consistent with the photo-z
(\S\ref{sec:host}).  The {\it HST} observations built up the infrared
(IR) light curve (\S\ref{sec:lightcurve}) and also provided a grism
spectrum of both the SN and its host galaxy (\S\ref{sec:grism}).

\section{Color}
\label{sec:color}

A key improvement for the CANDELS and CLASH programs over past {\it HST} 
SN surveys is the availability of the F350LP filter on the WFC3-UVIS
camera.  This very broad ``white light'' filter transmits all optical
light, and is therefore extremely efficient. In an exposure of just 400~s we 
can reach a signal-to-noise ratio (S/N) of 20 for point sources as faint as
Vega magnitude 25.4. For a SN at $z=1.5$ the F350LP filter samples the
rest-frame ultraviolet, blueward of 3600 \AA, offering a stark
contrast between mostly blue \CCSNe\ and much redder \SNeIa.

With the date of peak brightness well defined, the F350LP$-$F160W
color-magnitude measurement from the discovery epoch provides an
important early classification indicator, as shown in
Figure~\ref{fig:colormag}. Using the known redshift of the host galaxy
(\S\ref{sec:host}), we use the SNANA simulation
tools \citep{Kessler:2009a} to generate 10,000 simulated SNe of Types
Ia, Ib/c, and II.  The \CCSN\ simulation is limited by the dearth of
well-measured light curve templates at low redshift.  Whereas
the \SNIa\ simulation is defined by a parameterized model with strong
empirical constraints, our simulated \CCSNe\ are based on just 44
low-$z$ templates.  This deficiency is reflected in the patchy and
streaky \CCSN\ contours in both Figure~\ref{fig:colormag}
and \ref{fig:deltadelta}.  Nevertheless, the SNANA simulation provides
the most complete picture of the \CCSN\ population possible with
currently available data.

The observed F350LP$-$F160W color for \primo\ at peak brightness is
consistent with a Type Ia SN at $z=1.55$, but could also be matched by
\CCSN\ templates at that redshift. The F160W ($H$) magnitude
is more difficult to reconcile with a core-collapse model. One does
not want to use a strong prior on the apparent magnitude of high-$z$
SNe as the basis for classification, as cosmological effects could
have an impact on observed magnitudes.  However, \primo\ is observed
at a full 2~mag brighter than the brightest Type II SN of similar
color.  The less common Type Ib/c SNe come closer to matching \primo's
F160W magnitude, but only in the case of the very rare over-luminous
SNe\,Ib/c, which comprise less than 4\% of the total \CCSN\ population
at low redshift \citep{Arcavi:2010,Smith:2011}.  The rate of
over-luminous \CCSNe\ may be enhanced in low-metallicity
environments \citep[see][for a recent review]{Modjaz:2011}. This could
make these objects more common at high redshift, although estimates
suggest that they will still comprise just a few percent of the
total \CCSN\ population.

Thus, \primo's position on the color-magnitude diagram provides good
evidence for an initial classification as a \SNIa.  After making this
assessment, follow-up observations were then executed to fill out the
IR light curve and measure the SN spectrum, providing further
diagnostics to confirm or refute this classification.

\section{Host Galaxy}
\label{sec:host}

As shown in Figure~\ref{fig:discovery}(b), \primo\ appeared very close
to a relatively bright galaxy in the CANDELS IR imaging. As can be
seen in the very deep multi-band imaging of the HUDF in
Figure~\ref{fig:discovery}(a), this galaxy has a bright core on its
northern extreme, and a fainter arm extending southward.  With no
other detectable objects within 1\farcs5, we made an initial
assumption that this object was in fact \primo's host galaxy. However,
it is possible that the faint southern arm representing a foreground
or background galaxy. To examine this possibility, we measured
isophotal photometry, treating the two components of the galaxy as
separate objects.  The colors of the core are identical to the
southern arm to within the measurement error (0.04 mag) across seven
optical and IR bands from ACS and WFC3.  

Immediately after discovery, the photometric redshift (photo-$z$) of a
SN host galaxy is one of the most important tools for segregating targets of 
interest from impostor transients (\eg, active galactic nuclei, \CCSNe\ at 
lower redshifts, \etc).  The probability distribution of the
photo-$z$ for \primo's host is sharply peaked at $z=1.56$ by the extremely
deep existing {\it HST} data in the HUDF, with a 68\% confidence range of
$1.51 < z < 1.64$ and a 95\% confidence range $1.45 < z < 1.76$.  The
principal constraint comes from the offset between optical photometry
and IR photometry, since the 4000~\AA\ break falls between the optical
and IR at $z \approx 1.5$. This emphasizes the importance of accurate IR
photometry for determining the host galaxy photo-$z$ at $z>1.5$.

 A VLT observation on UT 2010 Oct16 using the X-Shooter spectrograph
targeted SN Primo and the bright host galaxy core in a single slit.
Both the SN and the host continuum were too faint for detection (the
host brightness is comparable to \primo\ at peak, with J=24.3 and
H=24.4 AB mag).  However, the X-shooter spectrum revealed 
strong emission lines from H$\alpha$, [O~III], and [O~II].  These
lines fixed the redshift of the host with exquisite precision to
z=1.54992$\pm$0.00007, and also confirmed the initial photometric
classification of the host as a strongly star-forming galaxy.  A
complete discussion of the VLT observations and a more detailed
discussion of the \primo\ host galaxy will be presented by Frederiksen
et al. (2012, in prep).

\section{Light Curve}
\label{sec:lightcurve}

\insertfigfloat{FIG/fig5}{\figlightcurve}

As detailed in Table~\ref{tab:lightcurve}, the fading light curve was
followed from 2010 November through 2011 January, with 6 visits from
the SN follow-up program and 2 return visits from CANDELS. The light
curve is plotted in Figure~\ref{fig:lightcurve} with the best-fitting
template (SN 2005cf), as determined by the SOFT light curve
classification program
\citep{Rodney:2009,Rodney:2010a}.  Reinforcing the color
classification, we find that this best-fitting light curve template is
a normal SNIa at the redshift of the host galaxy (SN 2005cf has a
light curve width of \dmfifteen$(B) = 1.05 \pm 0.03$ mag). The SOFT
fit with various \SNIa\ light curve templates can allow for
$0<\Av<0.5$ mag of host-galaxy extinction.  Using the MLCS2k2 light
curve fitter \citep{Jha:2007}, we find similarly undistinguished fit
parameters: a very normal light curve shape parameter
$\Delta=-0.12\pm0.10$ and a host-galaxy extinction of
$\Av=0.14\pm0.14$.  These estimates of low $\Av$ are supported by the
VLT spectrum of the host, in which the Balmer decrement of the
emission lines is consistent with no reddening (Frederiksen et
al. 2012, in prep).

The light curve shape measurement provides a strong test of the \SNIa\
classification, as shown in Figure~\ref{fig:deltadelta}.
The \dmfifteen\ parameter quantifies the shape of a \SNIa\ light curve as
the change in magnitude from peak to 15 rest-frame days past
maximum \citep{Phillips:1993}.  For \SNeIa\ this parameter varies over
a small range and is well correlated across adjacent photometric
bands.  Figure~\ref{fig:deltadelta} shows that \primo's light curve
shape is consistent with the narrow band defined by our simulated
\SNeIa\ at $z=1.55$.  

The patchy \CCSN\ contours in Figure~\ref{fig:deltadelta} do not
reflect the true range of \CCSN\ light curve shapes, due to the
incomplete information available for modeling this population (see
discussion in \S\ref{sec:color}).  Thus, one cannot rule out the
possibility that \primo's position in \dmfifteen\ parameter space is
consistent with a Type Ib/c population that is underrepresented by the
SNANA templates.  The available information, however, lends strong
support to the initial classification as a \SNIa.

\section{Grism Spectrum}
\label{sec:grism}

  \insertfigwide{FIG/fig6}{\figspectrum}

In the first epoch of follow-up observations, \primo\ was observed
using the WFC3 G141 grism (resolution $R \approx 130$, 1100--1700 nm).
Observations were executed in three visits spaced over 7
days,\footnote{The initial observing plan would have completed the
grism observations in a single day, but a guiding error resulted in a
failure during the second visit, so it had to be rescheduled for the
subsequent week.} for a total integration time of 21.7~ks (see
Table~\ref{tab:lightcurve}). \primo\ was separated from the nucleus of
its host galaxy by $\sim 0\farcs5$ (0\farcs23 E, 0\farcs44 S),
so it was possible to select the orientation to minimize contamination
of the SN spectrum by its own host. These grism data were processed
using the aXe software package,\footnote{\url{http://axe.stsci.edu}}
and spectra at the location of \primo\ and the bright core of its host
were separately extracted.  Both spectra were dereddened with a
correction for E(B-V)=0.008 mag of Milky Way extinction.

\subsection{ Host Galaxy Contamination}

The two-dimensional (2D) grism trace is shown in the bottom-left panel
of Figure~\ref{fig:spectrum}, in which the bright emission knots from
H$\beta$, [O~III], and H$\alpha$ are readily apparent.  The spectrum
extracted at the location of \primo\ is visible as the distinct upper
trace. Galaxy emission lines are also visible in this spectrum.  To
remove the host component from the \primo\ spectrum, we need an
estimate of the host emission features as well as the continuum at the
location of the SN.  The galaxy's emission lines may be unrelated to
the continuum strength, as the emission features are driven by the
galaxy's gas while the continuum largely reflects the stellar content.
Can the grism spectrum of the host core provide an adequate model for
the host light at \primo's location?

As noted in Section~\ref{sec:host}, the optical and NIR colors of the
host galaxy (derived from pre-explosion HUDF imaging) do not have an
observable gradient across the visible extent of the galaxy.  In
particular, the flux ratio from the core to the southern arm of the
galaxy is the same for the F125W band (dominated by strong H$\beta$
and [O~III] emission) as it is for continuum-dominated F105W and
optical bands, to within the measurement uncertainty.  This suggests
that both the emission and continuum components of the host spectrum
as measured at the core should be valid when applied to the southern
arm where \primo\ appears.

To adapt the spectrum of the host galaxy core for use in contamination
removal requires flux scaling, a wavelength shift and smoothing.  To
determine the appropriate scaling of the continuum flux, we first
measure the isophotal flux in F125W and F160W for the host galaxy core
and the southern arm, treating the two as independent objects.  The
ratio of NIR flux from core to arm is 2.5, which we adopt as our
continuum scaling factor.  For the emission line regions (demarcated
by gray bars in Figure~\ref{fig:spectrum}) we scale the core flux by a
factor of 2.0, which results in a good match from core to arm for the
[OIII]+H$\beta$ region.  

Next, we shift the host spectrum by $-32$~\AA\ in the rest frame.  As
shown in Figure~\ref{fig:discovery}, the dispersion axis of the grism
observations was roughly 90 degrees East of North, and \primo\ was
offset to the East of its host by 0\farcs23 along that axis.  In
the slitless grism observations, this spatial separation along the
dispersion direction is translated into a wavelength separation
between the SN features and the superimposed host-galaxy features. The
wavelength shift corrects for this offset.

Finally, we apply a broad smoothing filter ($\sim$250~\AA\ in the rest
frame) to the regions of the host spectrum between emission lines,
while leaving the lines untouched.  This smoothing avoids the
introduction of additional noise when this core spectrum is subtracted
from the SN.  The final host galaxy spectrum -- scaled, shifted and
smoothed -- is shown in the center-left panel of
Figure~\ref{fig:spectrum}).  Subtracting this from the spectrum taken
at the location of the SN yields the final host-subtracted SN spectrum
shown in the upper left panel.

As shown in the upper left panel of Figure~\ref{fig:spectrum}, the
host-subtracted SN spectrum is noisy and spans only $\sim$2000~\AA\ in
rest wavelength.  The noise in the SN spectrum is characterized by
adjacent positive and negative spikes with a width of 50-100 \AA\ (or
20-40 \AA\ in the rest frame). The dispersion of the G141 grism
is 46.5 \AA\,pix$^{-1}$, which we sub-sample to 21.5 \AA\,pix$^{-1}$.
Our analysis here is directed at identifying spectral features of SN
sub-classes that are much broader than this in the rest frame. Binning
the rest-frame spectrum into wavelength bins of width 80~\AA\ (black
points in Figure~\ref{fig:spectrum}) removes this high frequency noise
without obscuring broader features.

\subsection{ Spectral Confirmation }
\label{sec:confirmation}

The relatively weak signal in this spectrum is insufficient for a pure
spectral {\em classification}, but we argue below that it is enough to
provide a spectral {\em confirmation} of this object as a normal Type
Ia SN.  A true spectral classification would require that the object
be assigned to a SN (sub)class without any other information, or
perhaps with only weak priors on age and redshift.  In the case
of \primo, we have already built up a series of strong classification
indicators from the redshift, magnitude, color, and light curve shape.
Taken together, this evidence provides a prediction that the grism
data should show spectral features consistent with a \SNIa\ at about 6
days past maximum light.  We can confirm or refute this prediction by
testing for the presence of such features in the grism data.

The strongest test of spectroscopic confirmation should not rely on
any information derived from the broad-band indicators of
Sections~\ref{sec:color}-\ref{sec:lightcurve}.  For example, the grism
spectrum from HST is flux-calibrated, so we could require that a
matching spectral template be consistent with the absolute flux scale.
This, however, would essentially be re-using the magnitude
classification from Figure~\ref{fig:colormag}.  We already know that
an over-luminous Core Collapse SN would be required to match \primo's
photometry, so by working only in relative flux units we allow for the
possibility that any of the spectrally normal \CCSN\ represented in
our template library could be peculiarly over-luminous at $z=1.55$.
Similarly, the shape of the light curve is inconsistent with the broad
plateaus of Type II-P and some IIn SNe (Figures~\ref{fig:lightcurve}
and \ref{fig:deltadelta}. We could use this information to limit our
spectral template library to only Type I SNe. This would be re-using
the light curve shape classification indicator, so we include
all \CCSN\ varieties in the template matching test.

We do allow two pieces of prior information to inform the spectral
test. First, we use the redshift of $z=1.55$ derived from host galaxy
emission lines to define the rest wavelength scale of the SN spectrum
(i.e. we reject the possibility that \primo\ is a foreground or
background object not associated with the adjacent galaxy).  Second,
we define the age of the SN at the time of the grism observations as
6$\pm$3 days past maximum light.  This age is derived directly from
the observed photometry, independent of any light curve fitting.

\subsection{ Spectral Cross Correlation Test }
\label{sec:snid}

The final reduced spectrum suggests a series of absorption features
spaced by roughly 400 \AA\ in rest wavelength.  This is qualitatively
consistent with the expected shape of a Type Ia spectrum: the broad
trough from 4600 to 5100~\AA\ is roughly consistent with an Fe/Si
complex, and narrower drops around 5300, 5700 and 6100~\AA are in line
with S\,II, and Si\,II absorption features. These features are
characteristic of a normal \SNIa\ spectrum and can also be found in
SN\,Ib/c spectra shortly after maximum, although they are typically
less pronounced than in \SNeIa\
\citep{Filippenko:1997}.

For a more quantitative evaluation of the \primo\ spectrum, we use the
SuperNova IDentification program \citep[SNID,][]{Blondin:2007}, which
does a cross-correlation comparison of an input spectrum against a
library of templates.  SNID was designed to be insensitive to the
shape of the continuum (it fits the continuum with a high-order
polynomial and removes it), so that it can provide spectral
classifications based principally on emission and absorption features.
This satisfies our requirement from \S\ref{sec:confirmation} that the
spectral classification test with SNID is independent of the priors
derived from broad-band magnitudes in Figures~\ref{fig:colormag}.
Additionally, SNID's continuum removal makes it insensitive to any
reddening of the SN spectrum due to host galaxy extinction, so we do
not apply any host extinction corrections to \primo's spectrum.  

Due to the relatively weak signal, SNID is unable to find a compelling
match with any template spectrum. \citet{Blondin:2007} defines the
quality parameter $r$lap, which combines the degree of overlap with
the strength of the cross-correlation signal.  A good SNID fit
typically has $r$lap$>5$, but for \primo\ SNID finds no spectral
template match with $r$lap$>$3.7.  Although this is not sufficient for
a true spectral {\em classification}, we can still complete the
spectroscopic {\em confirmation} test by considering the relative
quality of fit for the best-fitting templates for each SN class.

The best match from the SNID template library is a normal \SNIa\ (SN
1996X) at 8 days past maximum, shown in the upper right panel of
Figure~\ref{fig:spectrum}.  The best fitting core collapse templates
from the Type Ib/c and Type II families of templates are shown in the
center and lower right panels.  To quantify the quality of the \SNIa\
fit relative to the \CCSN\ templates, we compute the $\chi^2$
statistic and measure the integrated tail probability (the $p$-value).
Due to the narrow restrictions on redshift and age, the only remaining
variable for SNID is the choice of template, so the $\chi^2$ test has
21 degrees of freedom. The Ia match gives $\chi^2/\nu=24/21=1.18$ for
a $p$-value of 0.24, meaning that the $\chi^2$ test cannot reject this
Ia template as an acceptable model for the observed data.  The top
Type Ib/c template has $\chi^2/\nu=35/21=1.64$ for a $p$-value of
0.03, rejecting the model with 97\% confidence.  The best Type II
model is significantly worse, giving $\chi^2/\nu=51/21=2.45$ or
$p$=0.0002, which rejects the model with 99.98\% confidence.

When comparing an observed SN to a set of spectral templates, one must
be careful that the makeup of the template library does not bias the
$\chi^2$ test in favor of a Type Ia classification. The number of
published Type Ia SN spectra in the literature is much larger than for
\CCSNe, so if the \CCSN\ population is not adequately represented in
the template library then the $\chi^2$ value for the best Ia match will
be better simply by virtue of having more templates to choose from.
For \primo, we have limited the SNID library to spectral templates
observed in the range $6\pm3$ days after peak.  This leaves SNID with 72
Type Ia spectra from 31 normal \SNeIa, compared to 74 CCSN spectra from
24 separate CC SNe. Thus, the SNID template library is well balanced
for an unbiased $\chi^2$ test.

As described above, the SNID template matching algorithm is
insensitive to the broad shape of the continuum, so the subtraction of
a heavily smoothed host continuum should not have a strong effect.  As
a check that the template matching is not driven by host subtraction,
we repeated the above procedure with no subtraction of the host
continuum.  The emission line regions must be excluded from the fit,
but the SN light still dominates the flux in the inter-line region.
The best SNID matches are even worse in $r$lap and $\chi^2$, but the
best-fitting spectral template is still a normal Type Ia SN.

Binning the spectrum to smooth over the high frequency noise makes it
easier to identify the broad absorption features that are
characteristic of most SNe.  However, this procedure could lead to a
classification bias by obscuring sharp features that would challenge
the \SNIa\ classification. To test for such bias, the fitting
procedure described above was repeated twice.  First, with an
alternate smoothing using a 10-pixel median filter, and then with no
smoothing at all.  Both of these approaches significantly degrade the
quality of the template match for all classes, but in both instances
the Type Ia template remains the best match.  The 80 \AA\ binning
depicted in Figure~\ref{fig:spectrum} provides the strongest
discrimination between classes, and unambiguously favors the Type Ia
template match.

\section{Discussion}
\label{sec:discussion}

The discovery and confirmation of \primo\ demonstrates the new
capability of {\it HST} to both detect and follow \SNeIa\ at redshifts above
1.5 using the WFC3 IR detector.  Given the depth of our survey
observations in the IR, we are able to detect a normal \SNIa\ like
\primo\ to redshifts as high as $z \approx 2.3$, where it
would appear with an unreddened peak (Vega) magnitude around
F125W = 25.3 and F160W = 25.1 mag with a S/N of 10 (\ie, uncertainties 
of 0.1 mag).

After detecting SNe at $z \approx 2$, however, the real challenge lies
in the follow-up campaign.  The classification of \primo\ as a \SNIa\
is built on four layers of evidence.  First, the F350LP$-$F160W color
and the host-galaxy photometric redshift suggest that the object may
be a high-$z$ \SNIa\ (Figure~\ref{fig:colormag}).  Second,
ground-based follow-up spectra from the VLT pin down the redshift,
narrowing the range of possible models and strengthening the \SNIa\
case. Third, in Figure~\ref{fig:deltadelta} a well-sampled light curve
defines \primo's position in the \dmfifteen\ parameter space as
consistent with the narrow \SNIa\ band.  Finally, the spectrum shown
in Figure~\ref{fig:spectrum} reveals \SNIa\ spectral features that
confirm the \SNIa\ classification.  Although this final test falls
short of a pure spectroscopic classification, our spectral
confirmation procedure is strengthened by keeping it largely
independent of the photometric classification tests: no spectral
templates were excluded by invoking the luminosity prior of
Figure~\ref{fig:colormag} or the light curve shape constraints of
Figure~\ref{fig:deltadelta}, but we still cannot find a better match
to \primo's spectral features with any \CCSN\ template.

As the CANDELS+CLASH SN survey progresses, the first question that the
growing $z>1.5$ SN sample will be able to address is simply, how many
\SNeIa\ are there at $z>1.5$?  This rate measurement provides an important
constraint on models of \SNIa\ progenitors through measurement of
the DTD.  Past {\it HST} programs using the ACS were able to extend the
\SNIa\ rate measurement to $z \approx 1.5$, but only with very weak
statistical constraints \citep[][D08]{Strolger:2004}.  The D08 results
suggest a decline in the \SNIa\ rate at $z>1.2$, but that claim rests
on only three SN detections that populate the highest redshift
bin. Ground-based rate measurements from the Subaru Deep Field (SDF)
survey have claimed a higher \SNIa\ rate at $z \approx 1.5$ (G11).  The SDF
results have larger statistical significance (their sample has
10 \SNeIa\ at $z \approx 1.5$, where D08 has 3).  However, there is a
greater potential for classification errors, as the G11 sample has
only single-epoch detections with a one-year interval, no
spectroscopic confirmation of the SNe, and no spectroscopic host
redshifts above $z=1.2$.

The CANDELS+CLASH survey will be able to resolve this dispute
observationally with two rate measurements.  First, in the $1<z<1.5$
range, this program is the first SN search to use IR bands for SN
discovery, making it less sensitive to dust extinction that could have
obscured SNe in optical surveys. Second, this survey will for the
first time extend the \SNIa\ rate measurement to $z \approx 2$.
Extrapolating the rate measurements from D08 and G11, we expect to
discover $\sim 10$ \SNeIa\ at $z>1.5$ over the three-year CANDELS+CLASH
program.  As with \primo, all of these objects will have secure
classifications and the best available redshifts.

Finally, the full sample of $z > 1.5$ SNe will enable a direct test
for evolution in the properties of \SNeIa.  In
Figure~\ref{fig:hubblepoints} we see
that \primo\ is consistent with the standard \LCDM\ model.
Figure~\ref{fig:lightcurve} shows that \primo\ is very well matched by
a light curve template based on SN 2005cf, a normal \SNIa\ with a
light curve width of \dmfifteen$(B) = 1.05$ mag.  The spectral features
shown in Figure~\ref{fig:spectrum} are also consistent with a
normal \SNIa.  Thus, \primo\ alone provides no evidence for evolution
of the \SNIa\ population with redshift, but with only a single
well-studied object so far one cannot draw any meaningful
conclusions. Extending these comparisons to the final sample of
$\sim 10$ high-$z$ \SNeIa\ may provide useful constraints on evolution of
the \SNIa\ population.

\bigskip
We would like to thank the anonymous referee for a helpful
discussion that significantly improved this paper.  Thanks also to our
Program Coordinators, Beth Perriello and Tricia Royle, our Contact
Scientist Sylvia Baggett, and the entire \Hubble\ planning team, for
their efforts in support of this challenging program. The WFC3 team
has made substantial contributions to the program by calibrating and
characterizing the instrument, and have provided much useful advice.
We thank David Koo and Asantha Cooray of CANDELS for helpful
discussions and review of this paper, and Marc Postman and Larry
Bradley of CLASH for their steadfast support of the CLASH SN search
program. Thanks to Zoltan Levay of STScI for assistance in image
preparation.

Financial support for this work was provided by NASA through grants
HST-GO-12060 and HST-GO-12099 from the Space Telescope Science
Institute, which is operated by Associated Universities for Research
in Astronomy, Inc., under NASA contract NAS 5-26555.  Support for this
research at Rutgers University was provided in part by NSF CAREER
award AST-0847157 to SWJ.  The Dark Cosmology Centre is supported by
the Danish National Research Foundation.

{\it Facilities:} \facility{HST (WFC3)} \facility{VLT (X-shooter)}
\smallskip

\bibliographystyle{apj}
\bibliography{bibdesk}

\end{document}

%% file: lightcurve.tex
\begin{small}
\begin{deluxetable}{lllrl} 
\tablecolumns{4}
\tablecaption{\primo\ Photometric Data\label{tab:lightcurve}}
 
\tablehead{ \colhead{UT\,Date}
	  & \colhead{MJD}
	  & \colhead{Filter}
	  & \colhead{Exp.Time}
	  & \colhead{Vega Mag}
	  }
\startdata
2010 Oct\,10.9$^*$  & 55479.9 & F350LP & 434  &  27.2$\pm$0.2  \vspace{2mm}\\
2010 Aug\,06.9      & 55414.9 & F125W  & 11023 &  $>$28.0         \\
2010 Oct\,10.9$^*$  & 55479.9 & F125W  & 1006  &  24.37$\pm$0.05  \\
2010 Nov\,01.4      & 55501.4 & F125W  & 406   &  24.48$\pm$0.06  \\
2010 Nov\,07.7      & 55507.7 & F125W  & 1306  &  24.57$\pm$0.03  \\
2010 Nov\,21.2      & 55521.2 & F125W  & 2512  &  24.94$\pm$0.04  \\
2010 Nov\,28.4      & 55528.4 & F125W  & 1006  &  25.03$\pm$0.07  \\
2010 Dec\,05.2      & 55535.2 & F125W  & 2512  &  25.62$\pm$0.07  \\
2010 Dec\,13.8      & 55543.8 & F125W  & 5123  &  25.83$\pm$0.06  \\
2011 Jan\,17.7      & 55578.7 & F125W  & 1006  &  27.00$\pm$0.40  \\
2011 Mar\,05.6      & 55625.6 & F125W  & 1006  &  $>$27.2         \vspace{2mm}\\
2010 Aug\,08.4      & 55416.4 & F160W & 11024 &  $>$27.0         \\ 
2010 Sep\,02.9      & 55441.9 & F160W & 10824 &  $>$26.2         \\
2010 Sep\,14.9      & 55453.9 & F160W & 5512  &  25.39$\pm$0.15  \\
2010 Oct\,10.9$^*$  & 55479.9 & F160W & 1056  &  23.98$\pm$0.04  \\
2010 Oct\,26.6      & 55495.6 & F160W & 1206  &  24.27$\pm$0.05  \\
2010 Nov\,07.7      & 55507.7 & F160W & 1306  &  24.53$\pm$0.08  \\
2010 Nov\,22.9      & 55522.9 & F160W & 2512  &  24.61$\pm$0.06  \\
2010 Nov\,28.4      & 55528.4 & F160W & 956   &  24.57$\pm$0.11  \\
2010 Dec\,05.3      & 55535.3 & F160W & 2512  &  24.88$\pm$0.14  \\
2011 Jan\,17.7      & 55578.7 & F160W & 1106  &  25.72$\pm$0.19  \\
2011 Mar\,05.6      & 55625.6 & F160W & 1106  &  $>$26.2         \vspace{2mm}\\
2010 Oct\,26.7      & 55495.7 &  G141 &  6618 & (grism obs) \\
2010 Nov\,01.5      & 55501.5 &  G141 & 15036 & (grism obs) \\
\enddata
\tablenotetext{*}{Date of discovery}
\end{deluxetable}
\end{small}

%% file: primo.bbl
\begin{thebibliography}{44}
\expandafter\ifx\csname natexlab\endcsname\relax\def\natexlab#1{#1}\fi

\bibitem[{{Arcavi} {et~al.}(2010){Arcavi}, {Gal-Yam}, {Kasliwal}, {Quimby},
  {Ofek}, {Kulkarni}, {Nugent}, {Cenko}, {Bloom}, {Sullivan}, {Howell},
  {Poznanski}, {Filippenko}, {Law}, {Hook}, {J{\"o}nsson}, {Blake}, {Cooke},
  {Dekany}, {Rahmer}, {Hale}, {Smith}, {Zolkower}, {Velur}, {Walters},
  {Henning}, {Bui}, {McKenna}, \& {Jacobsen}}]{Arcavi:2010}
{Arcavi}, I., {et~al.} 2010, \apj, 721, 777

\bibitem[{{Blondin} \& {Tonry}(2007)}]{Blondin:2007}
{Blondin}, S., \& {Tonry}, J.~L. 2007, \apj, 666, 1024

\bibitem[{{Caldwell} {et~al.}(1998){Caldwell}, {Dave}, \&
  {Steinhardt}}]{Caldwell:1998}
{Caldwell}, R.~R., {Dave}, R., \& {Steinhardt}, P.~J. 1998, Physical Review
  Letters, 80, 1582

\bibitem[{{Conley} {et~al.}(2011){Conley}, {Guy}, {Sullivan}, {Regnault},
  {Astier}, {Balland}, {Basa}, {Carlberg}, {Fouchez}, {Hardin}, {Hook},
  {Howell}, {Pain}, {Palanque-Delabrouille}, {Perrett}, {Pritchet}, {Rich},
  {Ruhlmann-Kleider}, {Balam}, {Baumont}, {Ellis}, {Fabbro}, {Fakhouri},
  {Fourmanoit}, {Gonz{\'a}lez-Gait{\'a}n}, {Graham}, {Hudson}, {Hsiao},
  {Kronborg}, {Lidman}, {Mourao}, {Neill}, {Perlmutter}, {Ripoche}, {Suzuki},
  \& {Walker}}]{Conley:2011}
{Conley}, A., {et~al.} 2011, \apjs, 192, 1

\bibitem[{{Contreras} {et~al.}(2010){Contreras}, {Hamuy}, {Phillips},
  {Folatelli}, {Suntzeff}, {Persson}, {Stritzinger}, {Boldt}, {Gonz{\'a}lez},
  {Krzeminski}, {Morrell}, {Roth}, {Salgado}, {Jos{\'e} Maureira}, {Burns},
  {Freedman}, {Madore}, {Murphy}, {Wyatt}, {Li}, \&
  {Filippenko}}]{Contreras:2010}
{Contreras}, C., {et~al.} 2010, \aj, 139, 519

\bibitem[{{Dahlen} {et~al.}(2008){Dahlen}, {Strolger}, \&
  {Riess}}]{Dahlen:2008}
{Dahlen}, T., {Strolger}, L.-G., \& {Riess}, A.~G. 2008, \apj, 681, 462

\bibitem[{{Dom{\'{\i}}nguez} {et~al.}(2001){Dom{\'{\i}}nguez}, {H{\"o}flich},
  \& {Straniero}}]{Dominguez:2001}
{Dom{\'{\i}}nguez}, I., {H{\"o}flich}, P., \& {Straniero}, O. 2001, \apj, 557,
  279

\bibitem[{{Drout} {et~al.}(2011){Drout}, {Soderberg}, {Gal-Yam}, {Cenko},
  {Fox}, {Leonard}, {Sand}, {Moon}, {Arcavi}, \& {Green}}]{Drout:2011}
{Drout}, M.~R., {et~al.} 2011, \apj, 741, 97

\bibitem[{{Filippenko}(1997)}]{Filippenko:1997}
{Filippenko}, A.~V. 1997, \araa, 35, 309

\bibitem[{{Garnavich} {et~al.}(1998){Garnavich}, {Jha}, {Challis},
  {Clocchiatti}, {Diercks}, {Filippenko}, {Gilliland}, {Hogan}, {Kirshner},
  {Leibundgut}, {Phillips}, {Reiss}, {Riess}, {Schmidt}, {Schommer}, {Smith},
  {Spyromilio}, {Stubbs}, {Suntzeff}, {Tonry}, \& {Carroll}}]{Garnavich:1998b}
{Garnavich}, P.~M., {et~al.} 1998, \apj, 509, 74

\bibitem[{{Graur} {et~al.}(2011){Graur}, {Poznanski}, {Maoz}, {Yasuda},
  {Totani}, {Fukugita}, {Filippenko}, {Foley}, {Silverman}, {Gal-Yam},
  {Horesh}, \& {Jannuzi}}]{Graur:2011}
{Graur}, O., {et~al.} 2011, \mnras, 417, 916

\bibitem[{{Grogin} {et~al.}(2011){Grogin}, {Kocevski}, {Faber}, {Ferguson},
  {Koekemoer}, {Riess}, {Acquaviva}, {Alexander}, {Almaini}, {Ashby}, {Barden},
  {Bell}, {Bournaud}, {Brown}, {Caputi}, {Casertano}, {Cassata}, {Castellano},
  {Challis}, {Chary}, {Cheung}, {Cirasuolo}, {Conselice}, {Roshan Cooray},
  {Croton}, {Daddi}, {Dahlen}, {Dav{\'e}}, {de Mello}, {Dekel}, {Dickinson},
  {Dolch}, {Donley}, {Dunlop}, {Dutton}, {Elbaz}, {Fazio}, {Filippenko},
  {Finkelstein}, {Fontana}, {Gardner}, {Garnavich}, {Gawiser}, {Giavalisco},
  {Grazian}, {Guo}, {Hathi}, {H{\"a}ussler}, {Hopkins}, {Huang}, {Huang},
  {Jha}, {Kartaltepe}, {Kirshner}, {Koo}, {Lai}, {Lee}, {Li}, {Lotz}, {Lucas},
  {Madau}, {McCarthy}, {McGrath}, {McIntosh}, {McLure}, {Mobasher},
  {Moustakas}, {Mozena}, {Nandra}, {Newman}, {Niemi}, {Noeske}, {Papovich},
  {Pentericci}, {Pope}, {Primack}, {Rajan}, {Ravindranath}, {Reddy}, {Renzini},
  {Rix}, {Robaina}, {Rodney}, {Rosario}, {Rosati}, {Salimbeni}, {Scarlata},
  {Siana}, {Simard}, {Smidt}, {Somerville}, {Spinrad}, {Straughn}, {Strolger},
  {Telford}, {Teplitz}, {Trump}, {van der Wel}, {Villforth}, {Wechsler},
  {Weiner}, {Wiklind}, {Wild}, {Wilson}, {Wuyts}, {Yan}, \&
  {Yun}}]{Grogin:2011}
{Grogin}, N.~A., {et~al.} 2011, \apjs, 197, 35

\bibitem[{Guy {et~al.}(2007)Guy, Astier, Baumont, Hardin, Pain, Regnault, Basa,
  Carlberg, Conley, Fabbro, Fouchez, Hook, Howell, Perrett, Pritchet, Rich,
  Sullivan, Antilogus, Aubourg, Bazin, Bronder, Filiol, Palanque-Delabrouille,
  Ripoche, \& Ruhlmann-Kleider}]{Guy:2007}
Guy, J., {et~al.} 2007, \aap, 466, 11

\bibitem[{{Hachisu} {et~al.}(1996){Hachisu}, {Kato}, \&
  {Nomoto}}]{Hachisu:1996}
{Hachisu}, I., {Kato}, M., \& {Nomoto}, K. 1996, \apjl, 470, L97+

\bibitem[{{Hicken} {et~al.}(2009){Hicken}, {Challis}, {Jha}, {Kirshner},
  {Matheson}, {Modjaz}, {Rest}, {Michael Wood-Vasey}, {Bakos}, {Barton},
  {Berlind}, {Bragg}, {Brice{\~n}o}, {Brown}, {Caldwell}, {Calkins}, {Cho},
  {Ciupik}, {Contreras}, {Dendy}, {Dosaj}, {Durham}, {Eriksen}, {Esquerdo},
  {Everett}, {Falco}, {Fernandez}, {Gaba}, {Garnavich}, {Graves}, {Green},
  {Groner}, {Hergenrother}, {Holman}, {Hradecky}, {Huchra}, {Hutchison},
  {Jerius}, {Jordan}, {Kilgard}, {Krauss}, {Luhman}, {Macri}, {Marrone},
  {McDowell}, {McIntosh}, {McNamara}, {Megeath}, {Mochejska}, {Munoz},
  {Muzerolle}, {Naranjo}, {Narayan}, {Pahre}, {Peters}, {Peterson}, {Rines},
  {Ripman}, {Roussanova}, {Schild}, {Sicilia-Aguilar}, {Sokoloski}, {Smalley},
  {Smith}, {Spahr}, {Stanek}, {Barmby}, {Blondin}, {Stubbs}, {Szentgyorgyi},
  {Torres}, {Vaz}, {Vikhlinin}, {Wang}, {Westover}, {Woods}, \&
  {Zhao}}]{Hicken:2009}
{Hicken}, M., {et~al.} 2009, \apj, 700, 331

\bibitem[{{Hoeflich} {et~al.}(1998){Hoeflich}, {Wheeler}, \&
  {Thielemann}}]{Hoeflich:1998}
{Hoeflich}, P., {Wheeler}, J.~C., \& {Thielemann}, F.~K. 1998, \apj, 495, 617

\bibitem[{{Jha} {et~al.}(2006){Jha}, {Kirshner}, {Challis}, {Garnavich},
  {Matheson}, {Soderberg}, {Graves}, {Hicken}, {Alves}, {Arce}, {Balog},
  {Barmby}, {Barton}, {Berlind}, {Bragg}, {Brice{\~n}o}, {Brown}, {Buckley},
  {Caldwell}, {Calkins}, {Carter}, {Concannon}, {Donnelly}, {Eriksen},
  {Fabricant}, {Falco}, {Fiore}, {Garcia}, {G{\'o}mez}, {Grogin}, {Groner},
  {Groot}, {Haisch}, {Hartmann}, {Hergenrother}, {Holman}, {Huchra},
  {Jayawardhana}, {Jerius}, {Kannappan}, {Kim}, {Kleyna}, {Kochanek},
  {Koranyi}, {Krockenberger}, {Lada}, {Luhman}, {Luu}, {Macri}, {Mader},
  {Mahdavi}, {Marengo}, {Marsden}, {McLeod}, {McNamara}, {Megeath}, {Moraru},
  {Mossman}, {Muench}, {Mu{\~n}oz}, {Muzerolle}, {Naranjo}, {Nelson-Patel},
  {Pahre}, {Patten}, {Peters}, {Peters}, {Raymond}, {Rines}, {Schild},
  {Sobczak}, {Spahr}, {Stauffer}, {Stefanik}, {Szentgyorgyi}, {Tollestrup},
  {V{\"a}is{\"a}nen}, {Vikhlinin}, {Wang}, {Willner}, {Wolk}, {Zajac}, {Zhao},
  \& {Stanek}}]{Jha:2006}
{Jha}, S., {et~al.} 2006, \aj, 131, 527

\bibitem[{Jha {et~al.}(2007)Jha, Riess, \& Kirshner}]{Jha:2007}
Jha, S., Riess, A.~G., \& Kirshner, R.~P. 2007, \apj, 659, 122

\bibitem[{{Kessler} {et~al.}(2009{\natexlab{a}}){Kessler}, {Becker}, {Cinabro},
  {Vanderplas}, {Frieman}, {Marriner}, {Davis}, {Dilday}, {Holtzman}, {Jha},
  {Lampeitl}, {Sako}, {Smith}, {Zheng}, {Nichol}, {Bassett}, {Bender}, {Depoy},
  {Doi}, {Elson}, {Filippenko}, {Foley}, {Garnavich}, {Hopp}, {Ihara},
  {Ketzeback}, {Kollatschny}, {Konishi}, {Marshall}, {McMillan}, {Miknaitis},
  {Morokuma}, {M{\"o}rtsell}, {Pan}, {Prieto}, {Richmond}, {Riess}, {Romani},
  {Schneider}, {Sollerman}, {Takanashi}, {Tokita}, {van der Heyden}, {Wheeler},
  {Yasuda}, \& {York}}]{Kessler:2009}
{Kessler}, R., {et~al.} 2009{\natexlab{a}}, \apjs, 185, 32

\bibitem[{{Kessler} {et~al.}(2009{\natexlab{b}}){Kessler}, {Bernstein},
  {Cinabro}, {Dilday}, {Frieman}, {Jha}, {Kuhlmann}, {Miknaitis}, {Sako},
  {Taylor}, \& {Vanderplas}}]{Kessler:2009a}
---. 2009{\natexlab{b}}, \pasp, 121, 1028

\bibitem[{{Kiewe} {et~al.}(2012){Kiewe}, {Gal-Yam}, {Arcavi}, {Leonard},
  {Emilio Enriquez}, {Cenko}, {Fox}, {Moon}, {Sand}, {Soderberg}, \&
  {CCCP}}]{Kiewe:2012}
{Kiewe}, M., {et~al.} 2012, \apj, 744, 10

\bibitem[{{Koekemoer} {et~al.}(2011){Koekemoer}, {Faber}, {Ferguson}, {Grogin},
  {Kocevski}, {Koo}, {Lai}, {Lotz}, {Lucas}, {McGrath}, {Ogaz}, {Rajan},
  {Riess}, {Rodney}, {Strolger}, {Casertano}, {Castellano}, {Dahlen},
  {Dickinson}, {Dolch}, {Fontana}, {Giavalisco}, {Grazian}, {Guo}, {Hathi},
  {Huang}, {van der Wel}, {Yan}, {Acquaviva}, {Alexander}, {Almaini}, {Ashby},
  {Barden}, {Bell}, {Bournaud}, {Brown}, {Caputi}, {Cassata}, {Challis},
  {Chary}, {Cheung}, {Cirasuolo}, {Conselice}, {Roshan Cooray}, {Croton},
  {Daddi}, {Dav{\'e}}, {de Mello}, {de Ravel}, {Dekel}, {Donley}, {Dunlop},
  {Dutton}, {Elbaz}, {Fazio}, {Filippenko}, {Finkelstein}, {Frazer}, {Gardner},
  {Garnavich}, {Gawiser}, {Gruetzbauch}, {Hartley}, {H{\"a}ussler},
  {Herrington}, {Hopkins}, {Huang}, {Jha}, {Johnson}, {Kartaltepe},
  {Khostovan}, {Kirshner}, {Lani}, {Lee}, {Li}, {Madau}, {McCarthy},
  {McIntosh}, {McLure}, {McPartland}, {Mobasher}, {Moreira}, {Mortlock},
  {Moustakas}, {Mozena}, {Nandra}, {Newman}, {Nielsen}, {Niemi}, {Noeske},
  {Papovich}, {Pentericci}, {Pope}, {Primack}, {Ravindranath}, {Reddy},
  {Renzini}, {Rix}, {Robaina}, {Rosario}, {Rosati}, {Salimbeni}, {Scarlata},
  {Siana}, {Simard}, {Smidt}, {Snyder}, {Somerville}, {Spinrad}, {Straughn},
  {Telford}, {Teplitz}, {Trump}, {Vargas}, {Villforth}, {Wagner}, {Wandro},
  {Wechsler}, {Weiner}, {Wiklind}, {Wild}, {Wilson}, {Wuyts}, \&
  {Yun}}]{Koekemoer:2011}
{Koekemoer}, A.~M., {et~al.} 2011, \apjs, 197, 36

\bibitem[{{Komatsu} {et~al.}(2011){Komatsu}, {Smith}, {Dunkley}, {Bennett},
  {Gold}, {Hinshaw}, {Jarosik}, {Larson}, {Nolta}, {Page}, {Spergel},
  {Halpern}, {Hill}, {Kogut}, {Limon}, {Meyer}, {Odegard}, {Tucker}, {Weiland},
  {Wollack}, \& {Wright}}]{Komatsu:2011}
{Komatsu}, E., {et~al.} 2011, \apjs, 192, 18

\bibitem[{{Kuznetsova} {et~al.}(2008){Kuznetsova}, {Barbary}, {Connolly},
  {Kim}, {Pain}, {Roe}, {Aldering}, {Amanullah}, {Dawson}, {Doi}, {Fadeyev},
  {Fruchter}, {Gibbons}, {Goldhaber}, {Goobar}, {Gude}, {Knop}, {Kowalski},
  {Lidman}, {Morokuma}, {Meyers}, {Perlmutter}, {Rubin}, {Schlegel},
  {Spadafora}, {Stanishev}, {Strovink}, {Suzuki}, {Wang}, \&
  {Yasuda}}]{Kuznetsova:2008}
{Kuznetsova}, N., {et~al.} 2008, \apj, 673, 981

\bibitem[{{Li} {et~al.}(2011){Li}, {Leaman}, {Chornock}, {Filippenko},
  {Poznanski}, {Ganeshalingam}, {Wang}, {Modjaz}, {Jha}, {Foley}, \&
  {Smith}}]{Li:2011}
{Li}, W., {et~al.} 2011, \mnras, 412, 1441

\bibitem[{{Modjaz}(2011)}]{Modjaz:2011}
{Modjaz}, M. 2011, Astronomische Nachrichten, 332, 434

\bibitem[{{Perlmutter} {et~al.}(1999){Perlmutter}, {Aldering}, {Goldhaber},
  {Knop}, {Nugent}, {Castro}, {Deustua}, {Fabbro}, {Goobar}, {Groom}, {Hook},
  {Kim}, {Kim}, {Lee}, {Nunes}, {Pain}, {Pennypacker}, {Quimby}, {Lidman},
  {Ellis}, {Irwin}, {McMahon}, {Ruiz-Lapuente}, {Walton}, {Schaefer}, {Boyle},
  {Filippenko}, {Matheson}, {Fruchter}, {Panagia}, {Newberg}, {Couch}, \& {The
  Supernova Cosmology Project}}]{Perlmutter:1999}
{Perlmutter}, S., {et~al.} 1999, \apj, 517, 565

\bibitem[{Phillips(1993)}]{Phillips:1993}
Phillips, M.~M. 1993, \apj, 413, L105

\bibitem[{{Postman} {et~al.}(2011){Postman}, {Coe}, {Benitez}, {Bradley},
  {Broadhurst}, {Donahue}, {Ford}, {Graur}, {Graves}, {Jouvel}, {Koekemoer},
  {Lemze}, {Medezinski}, {Molino}, {Moustakas}, {Ogaz}, {Riess}, {Rodney},
  {Rosati}, {Umetsu}, {Zheng}, {Zitrin}, {Bartelmann}, {Bouwens}, {Host},
  {Infante}, {Jha}, {Jimenez-Teja}, {Kelson}, {Lahav}, {Lazkoz}, {Maoz},
  {McCully}, {Melchior}, {Meneghetti}, {Merten}, {Nonino}, {Patel}, {Regos},
  {Seitz}, {Sayers}, {Golwala}, \& {Van der Wel}}]{Postman:2011}
{Postman}, M., {et~al.} 2011, ArXiv e-prints

\bibitem[{{Richardson} {et~al.}(2002){Richardson}, {Branch}, {Casebeer},
  {Millard}, {Thomas}, \& {Baron}}]{Richardson:2002}
{Richardson}, D., {Branch}, D., {Casebeer}, D., {Millard}, J., {Thomas}, R.~C.,
  \& {Baron}, E. 2002, \aj, 123, 745

\bibitem[{{Riess} {et~al.}(1998){Riess}, {Filippenko}, {Challis},
  {Clocchiatti}, {Diercks}, {Garnavich}, {Gilliland}, {Hogan}, {Jha},
  {Kirshner}, {Leibundgut}, {Phillips}, {Reiss}, {Schmidt}, {Schommer},
  {Smith}, {Spyromilio}, {Stubbs}, {Suntzeff}, \& {Tonry}}]{Riess:1998}
{Riess}, A.~G., {et~al.} 1998, \aj, 116, 1009

\bibitem[{{Riess} \& {Livio}(2006)}]{Riess:2006}
{Riess}, A.~G., \& {Livio}, M. 2006, \apj, 648, 884

\bibitem[{{Riess} {et~al.}(2001){Riess}, {Nugent}, {Gilliland}, {Schmidt},
  {Tonry}, {Dickinson}, {Thompson}, {Budav{\'a}ri}, {Casertano}, {Evans},
  {Filippenko}, {Livio}, {Sanders}, {Shapley}, {Spinrad}, {Steidel}, {Stern},
  {Surace}, \& {Veilleux}}]{Riess:2001}
{Riess}, A.~G., {et~al.} 2001, \apj, 560, 49

\bibitem[{{Riess} {et~al.}(2007){Riess}, {Strolger}, {Casertano}, {Ferguson},
  {Mobasher}, {Gold}, {Challis}, {Filippenko}, {Jha}, {Li}, {Tonry}, {Foley},
  {Kirshner}, {Dickinson}, {MacDonald}, {Eisenstein}, {Livio}, {Younger}, {Xu},
  {Dahl{\'e}n}, \& {Stern}}]{Riess:2007}
---. 2007, \apj, 659, 98

\bibitem[{Riess {et~al.}(2004)Riess, Strolger, Tonry, Casertano, Ferguson,
  Mobasher, Challis, Filippenko, Jha, Li, Chornock, Kirshner, Leibundgut,
  Dickinson, Livio, Giavalisco, Steidel, Ben�tez, \& Tsvetanov}]{Riess:2004a}
Riess, A.~G., {et~al.} 2004, \apj, 607, 665

\bibitem[{{Rodney} \& {Tonry}(2009)}]{Rodney:2009}
{Rodney}, S.~A., \& {Tonry}, J.~L. 2009, \apj, 707, 1064

\bibitem[{{Rodney} \& {Tonry}(2010)}]{Rodney:2010a}
---. 2010, \apj, 715, 323

\bibitem[{{Smith} {et~al.}(2011){Smith}, {Li}, {Filippenko}, \&
  {Chornock}}]{Smith:2011}
{Smith}, N., {Li}, W., {Filippenko}, A.~V., \& {Chornock}, R. 2011, \mnras,
  412, 1522

\bibitem[{Strolger {et~al.}(2004)Strolger, Riess, Dahlen, Livio, Panagia,
  Challis, Tonry, Filippenko, Chornock, Ferguson, Koekemoer, Mobasher,
  Dickinson, Giavalisco, Casertano, Hook, Blondin, Leibundgut, Nonino, Rosati,
  Spinrad, Steidel, Stern, Garnavich, Matheson, Grogin, Hornschemeier,
  Kretchmer, Laidler, Lee, Lucas, de~Mello, Moustakas, Ravindranath,
  Richardson, \& Taylor}]{Strolger:2004}
Strolger, L.-G., {et~al.} 2004, \apj, 613, 200

\bibitem[{{Sullivan} {et~al.}(2011){Sullivan}, {Guy}, {Conley}, {Regnault},
  {Astier}, {Balland}, {Basa}, {Carlberg}, {Fouchez}, {Hardin}, {Hook},
  {Howell}, {Pain}, {Palanque-Delabrouille}, {Perrett}, {Pritchet}, {Rich},
  {Ruhlmann-Kleider}, {Balam}, {Baumont}, {Ellis}, {Fabbro}, {Fakhouri},
  {Fourmanoit}, {Gonz{\'a}lez-Gait{\'a}n}, {Graham}, {Hudson}, {Hsiao},
  {Kronborg}, {Lidman}, {Mourao}, {Neill}, {Perlmutter}, {Ripoche}, {Suzuki},
  \& {Walker}}]{Sullivan:2011}
{Sullivan}, M., {et~al.} 2011, \apj, 737, 102

\bibitem[{{Suzuki} {et~al.}(2011){Suzuki}, {Rubin}, {Lidman}, {Aldering},
  {Amanullah}, {Barbary}, {Barrientos}, {Botyanszki}, {Brodwin}, {Connolly},
  {Dawson}, {Dey}, {Doi}, {Donahue}, {Deustua}, {Eisenhardt}, {Ellingson},
  {Faccioli}, {Fadeyev}, {Fakhouri}, {Fruchter}, {Gilbank}, {Gladders},
  {Goldhaber}, {Gonzalez}, {Goobar}, {Gude}, {Hattori}, {Hoekstra}, {Hsiao},
  {Huang}, {Ihara}, {Jee}, {Johnston}, {Kashikawa}, {Koester}, {Konishi},
  {Kowalski}, {Linder}, {Lubin}, {Melbourne}, {Meyers}, {Morokuma}, {Munshi},
  {Mullis}, {Oda}, {Panagia}, {Perlmutter}, {Postman}, {Pritchard}, {Rhodes},
  {Ripoche}, {Rosati}, {Schlegel}, {Spadafora}, {Stanford}, {Stanishev},
  {Stern}, {Strovink}, {Takanashi}, {Tokita}, {Wagner}, {Wang}, {Yasuda}, \&
  {Yee}}]{Suzuki:2011}
{Suzuki}, N., {et~al.} 2011, ArXiv e-prints

\bibitem[{{Turner} \& {White}(1997)}]{Turner:1997}
{Turner}, M.~S., \& {White}, M. 1997, \prd, 56, 4439

\bibitem[{{Wang} {et~al.}(2009){Wang}, {Li}, {Filippenko}, {Foley}, {Kirshner},
  {Modjaz}, {Bloom}, {Brown}, {Carter}, {Friedman}, {Gal-Yam}, {Ganeshalingam},
  {Hicken}, {Krisciunas}, {Milne}, {Silverman}, {Suntzeff}, {Wood-Vasey},
  {Cenko}, {Challis}, {Fox}, {Kirkman}, {Li}, {Li}, {Malkan}, {Moore},
  {Reitzel}, {Rich}, {Serduke}, {Shang}, {Steele}, {Swift}, {Tao}, {Wong}, \&
  {Zhang}}]{WangX:2009}
{Wang}, X., {et~al.} 2009, \apj, 697, 380

\bibitem[{Wood-Vasey {et~al.}(2007)Wood-Vasey, Miknaitis, Stubbs, Jha, Riess,
  Garnavich, Kirshner, Aguilera, Becker, Blackman, Blondin, Challis,
  Clocchiatti, Conley, Covarrubias, Davis, Filippenko, Foley, Garg, Hicken,
  Krisciunas, Leibundgut, Li, Matheson, Miceli, Narayan, Pignata, Prieto, Rest,
  Salvo, Schmidt, Smith, Sollerman, Spyromilio, Tonry, Suntzeff, \&
  Zenteno}]{Wood-Vasey:2007}
Wood-Vasey, W.~M., {et~al.} 2007, \apj, 666, 694

\end{thebibliography}
